\documentclass[twocolumn,showpacs,preprintnumbers,amsmath,amssymb]{revtex4}

\usepackage{graphicx}
\usepackage{dcolumn}
\usepackage{bm}
\usepackage{soul}
\usepackage{color}

\begin{document}

\title{Interaction of a graphene sheet with a ferromagnetic metal plate}

\author{ Anh D. Phan$^{1}$, N. A. Viet$^{2}$, Nikolai A. Poklonski$^{3}$, Lilia M. Woods$^{1}$, Chi H. Le$^{4}$}
\affiliation{$^{1}$Department of Physics, University of South Florida, Tampa, Florida 33620, USA}%
\affiliation{$^{2}$Institute of Physics, 10 Daotan, Badinh, Hanoi, Vietnam}%
\affiliation{$^{3}$Physics Department, Belarusian State University, Minsk 220030, Belarus}%
\affiliation{$^{4}$School of Engineering, University of Greenwich, Medway, United Kingdom}%

\date{\today}

\begin{abstract}
Nanoscale surface forces such as Casimir and the van der Waals forces can have a significant influence on fabrication, handling and assembly processes as well as the performance of micro and nano devices. In this paper, the investigation and the calculation of the Casimir  force  between  a graphene sheet  and  a  ferromagnetic metal  substrate in  a  vacuum  are presented. The reflection coefficients of graphene are graphene-conductivity dependent, and the conductivity of graphene is described by the Kubo formalism. There is an effect of magnetic properties of the metal on the Casimir interaction. The magnetic effect plays a significant role at low temperatures or high value of chemical potential. The numerical results also demonstrate that the thickness of a metal slab has a minor influence on the Casimir force. The investigation and findings about the Casimir force in this study would lead to useful information and effective solutions for design and manufacturing of micro and nano devices, especially in the areas of micro and nano machining, fabrication, manipulation, assembly and metrology.

\end{abstract}

\maketitle

\section{Introduction}
Micro and nano technologies are potential economic engines and are considered as enabling technologies with exceptional economic capabilities. There have been increasingly more products in which the integrated micro and nano-electromechanical systems (MEMS/NEMS) are crucial in today's market because they allow for improved functionality, lower costs and higher quality. With the decrease in the size of MEMS and NEMS, additional nanoscale surface forces, such as the Casimir force and the van der Waals force, should be considered \cite{1,2}, especially in the areas of micro and nano machining, fabrication, manipulation, assembly and metrology. Therefore, a fundamental understanding of the Casimir force has recently been under intense discussion in the research and technological development (RTD) communities. 

The Casimir force can cause the small elements in a device to stick together. At small scales, the nanoscale surface forces may overcome elastic restoring actions in the device and lead to the plates' sticking during the fabrication process \cite{1}. It has been recently shown that Casimir forces may hamper the functioning of MEMS and NEMS devices by providing a pull-in instability \cite{36}. The Casimir force between two objects can induce adhesion and stiction leading to failures in devices \cite{2,3,32,36}. Particularly, when the size of a material is under the threshold length, the influence of the loading interaction on design and manufacturing of micro and nanodevices becomes much more prominent.

The Casimir interaction originates from the quantum electromagnetic fluctuations between two objects \cite{2,3,40}. It provides a fundamental understanding about nanoscience and has a significant influence on the high performance of devices. Over six decades, the Lifshitz theory has been employed to investigate this force in metal-metal \cite{4,5}, semiconductor-semiconductor \cite{6}, metal-superconductor \cite{7,28}, metamaterials-metamaterials \cite{8,9} and graphene-graphene systems \cite{10,11,12,34}. Results of some configurations have been experimentally examined and have good agreement with theoretical calculations \cite{3,13}.

The interaction between a non-magnetic metal with a magneto-dielectric material has also recently been discussed \cite{29,14,15}. These studies illustrate that the magnetic properties can significantly influence in the Casimir pressure. Theoretical and experimental studies also show that it is possible to obtain the repulsive Casimir force in such systems \cite{2,15,16}, and the results could provide possible solutions to handle the stiction and adhesion problems found in MEMS/NEMS devices. In terms of theory, the calculations depend substantially on the model that describes the dielectric function of metals. 

Graphene, an atomically thin layered material with novel properties, has gained a great deal of attention since its discovery \cite{30}. Researchers from a wide range of scientific fields have spent considerable effort in exploring its nature and its potential for practical applications \cite{17,18,19,20}. Recently obtained results demonstrate that graphene can be a promising candidate for next-generation electronic devices \cite{31}. The functions of graphene have been applied to diverse areas from biology to material science.
\begin{figure}[htp]
\includegraphics[width=7.5cm]{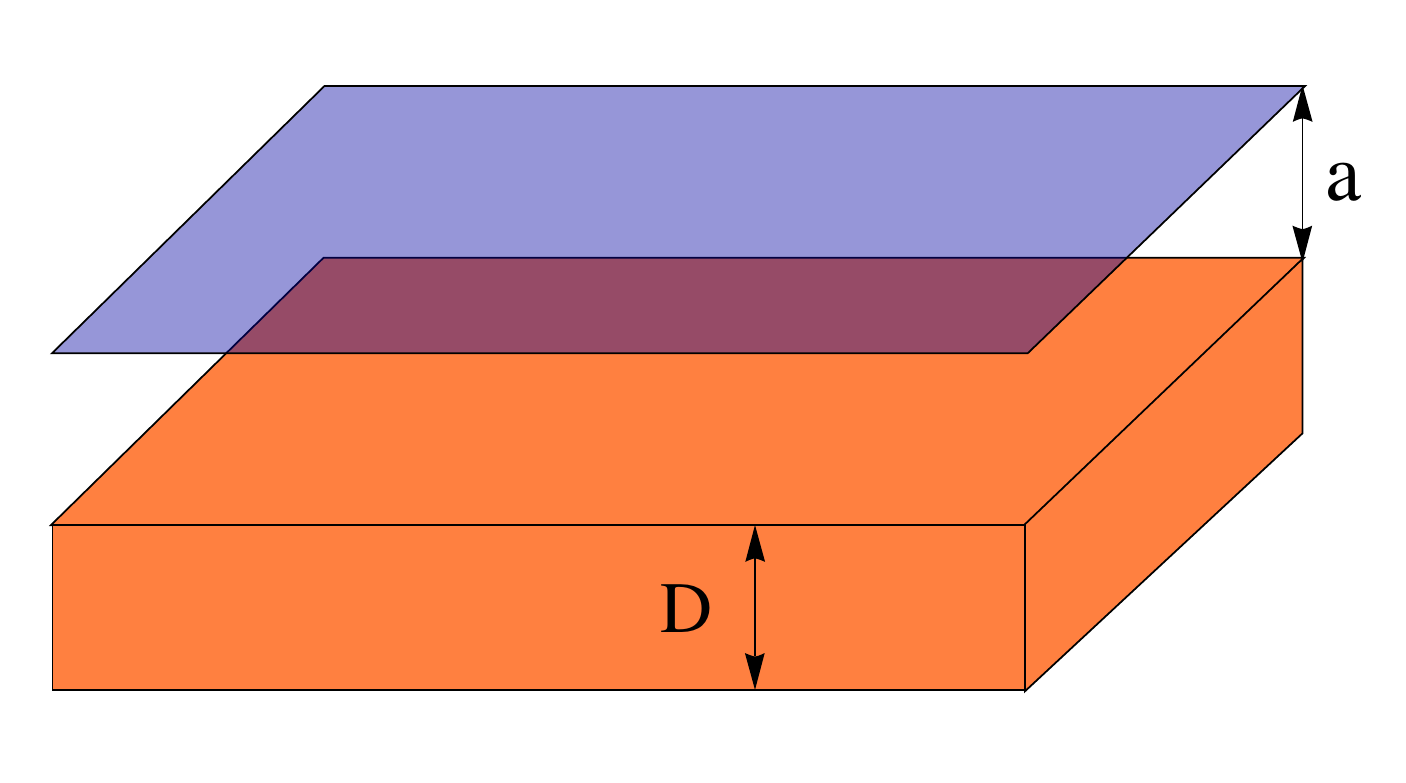}
\caption{\label{fig:0}(Color online) Schematics of a two-dimensional layer (graphene) and a substrate (FM metal) with thickness $D$.}
\end{figure}

In this paper, we calculate the Casimir force between a graphene layer and a ferromagnetic (FM) substrate shown in Fig.\ref{fig:0}. The FM material is taken to be Fe described via a dielectric and magnetic response properties. We utilize the Lifshitz theory to examine the role of the separation, temperature, and thickness of the substrate. We also investigate the influence of the magnetic response of the FM and show that in most cases its contribution is relatively small compared to the one from the dielectric response.

This paper is organized as follows: Sec. II presents the detailed expressions and calculations of the thermal Casimir interactions. The results and discussions are mentioned in Sec. III. Finally, Sec. IV presents the conclusions.
\section{CASIMIR INTERACTION BETWEEN GRAPHENE AND A METAL SUBSTRATE}
Here we calculate the Casimir force between the graphene sheet and a substrate in vacuum at a separation $a$. The force per unit area at temperature $T$ is given as \cite{3,21,34,24}

\begin{eqnarray}
& & F(a,T)=-\frac{k_{B}T}{\pi}\sum_{l=0}^{\infty}\left(1-\frac{1}{2}\delta_{l0}\right)\int_{0}^{\infty}q_{l}k_{\perp}dk_{\perp}\nonumber \\
& &\times \left(\frac{r_{TE}^{(1)}r_{TE}^{(2)}}{e^{2q_{l}a}-r_{TE}^{(1)}r_{TE}^{(2)}}+\dfrac{r_{TM}^{(1)}r_{TM}^{(2)}}{e^{2q_{l}a}-r_{TM}^{(1)}r_{TM}^{(2)}}\right),
\label{eq:1}
\end{eqnarray}
where $k_{B}$ is the Plank constant, $r_{TM}^{(1,2)}$ and $r_{TE}^{(1,2)}$ are the reflection coefficients corresponding to the transverse magnetic (TM) and transverse electric (TE) field modes. 

The reflection coefficients for the substrate are given as \cite{2,14,15}
\begin{eqnarray}
r_{TE}^{(1)} \equiv r_{TE}^{(1)}(i\xi_{l},k_{\perp})=\frac{\mu_{1}(i\xi_{l})q_{l}-k_{1}}{\mu_{1}(i\xi_{l})q_{l}+k_{1}},\nonumber \\
r_{TM}^{(1)} \equiv r_{TM}^{(1)}(i\xi_{l},k_{\perp})=\frac{\varepsilon_{1}(i\xi_{l})q_{l}-k_{1}}{\varepsilon_{1}(i\xi_{l})q_{l}+k_{1}},
\label{eq:2}
\end{eqnarray}
where
\begin{eqnarray}
q_{l} \equiv q_{l}(i\xi_{l},k_{\perp})=\sqrt{k_{\perp}^2+\frac{\xi_{l}^2}{c^2}},\nonumber \\
k_{1} \equiv k_{1}(i\xi_{l},k_{\perp})=\sqrt{k_{\perp}^2+\mu_{1}(i\xi_{l})\varepsilon_{1}(i\xi_{l})\frac{\xi_{l}^2}{c^2}}.
\label{eq:3}
\end{eqnarray}

It is important to note that $k_{\perp}$ is the wave vector component perpendicular to the plate, $c$ is the speed of light, $\xi_{l}=2\pi k_{B}Tl/\hbar$ is the Matsubara frequencies. The response properties of the metal are characterized by the dielectric function $\varepsilon_1(i\xi_l)$ and the permeability function $\mu_1(i\xi_l)$ which are frequency dependent along the imaginary axis ($\omega_l=i\xi_l$). We use the Drude model $\varepsilon_{1D}$ and the plasma model $\varepsilon_{1P}$ to describe the dielectric function of metals \cite{3,14,29}
\begin{eqnarray}
\varepsilon_{1D}(i\xi)=1+\frac{\omega_{p}^2}{\xi(\xi+\gamma_{p})},\nonumber \\
\varepsilon_{1P}(i\xi)=1+\frac{\omega_{p}^2}{\xi^2},
\label{eq:4}
\end{eqnarray}
where $\omega_{p}$ is the plasma frequency and the damping parameter is $\gamma_p$. At room temperature, the $l=0$ term of the Casimir-Lifshitz force in the metallic system is dominant. This reason allows us to consider the contribution of the static magnetic permeability $\mu(0) \gg 1$ on the Casimir force. For higher orders of $l$, one should put $\mu(\xi_l)=1$ \cite{29}. Both models lead to $r_{TM}^{(1)}(0)=1$. However, we can obtain two different expressions of $r_{TE}^{(1)}(0)$ when applying these two models \cite{29}
\begin{eqnarray}
r_{TE,D}^{(1)}(0,k_{\perp})=\frac{\mu(0)-1}{\mu(0)+1}, \nonumber\\
r_{TE,P}^{(1)}(0,k_{\perp})=\frac{\mu(0)ck_{\perp}-\sqrt{c^2k_{\perp}^2+\mu(0)\omega_p^2}}{\mu(0)ck_{\perp}+\sqrt{c^2k_{\perp}^2+\mu(0)\omega_p^2}},
\label{eq:8}
\end{eqnarray}
here $r_{TE,D}^{(1)}(0,k_{\perp})$ and $r_{TE,P}^{(1)}(0,k_{\perp})$ are the TE reflection coefficients corresponding to the Drude and plasma model, respectively.

For graphene, the reflection coefficients are found to be \cite{11,16}
\begin{eqnarray}
r_{TE}^{(2)} \equiv r_{TE}^{(2)}(i\xi_{l},k_{\perp})=-\frac{2\pi\xi_l\sigma/q_lc^2}{1+2\pi\xi_l\sigma/q_lc^2},\nonumber \\
r_{TM}^{(2)} \equiv r_{TM}^{(2)}(i\xi_{l},k_{\perp})=\frac{2\pi\sigma q_{l}/\xi_l}{1+2\pi\sigma q_{l}/\xi_l},
\label{eq:5}
\end{eqnarray}
where $\sigma\equiv \sigma(i\xi)$ is the 2D conductivity described via the the Kubo formalism \cite{11}
\begin{eqnarray}
\sigma(i\xi)=\frac{2e^2k_{B}T\ln(2)}{\pi\hbar^2\xi}+\frac{e^2\xi}{8\pi k_{B}T}\int_{0}^{\infty}\frac{\tanh(x)dx}{x^2+(\frac{\hbar\xi}{4k_{B}T})^2}.
\label{eq:6}
\end{eqnarray}

The first term in Eq.(\ref{eq:6}) corresponds to the intraband contribution, while the second term corresponds to the interband contribution. At low temperatures, the graphene conductivity $\sigma(i\xi_l)$ approaches to the universal value $\sigma_{0}=e^2/4\hbar$. 
\begin{figure}[htp]
\includegraphics[width=9cm]{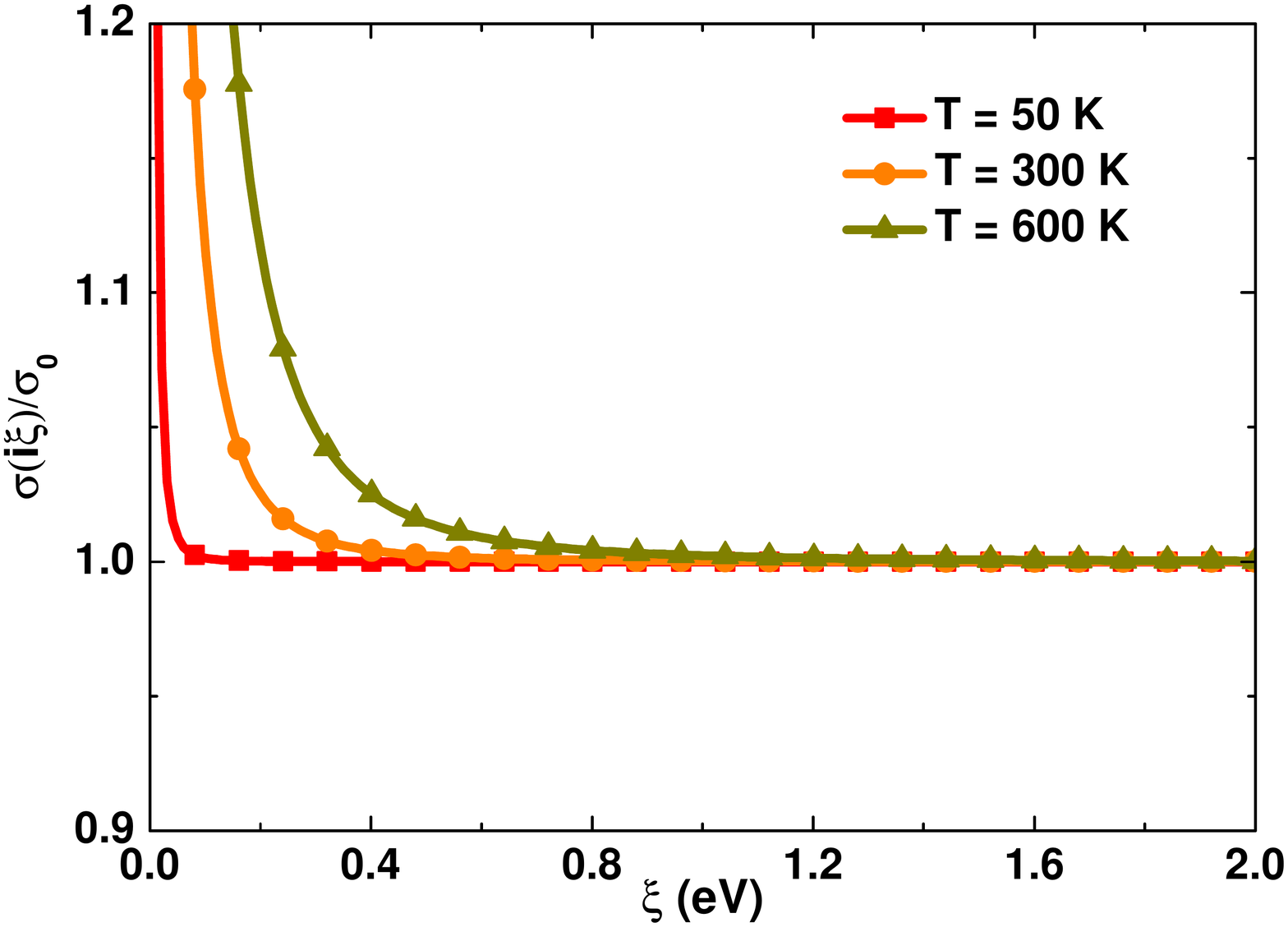}
\caption{\label{fig:6}(Color online) The normalized graphene conductivity $\sigma(i\xi)/\sigma_0$ vs frequency at given temperatures.}
\end{figure} 

Using Eq.(\ref{eq:5}) and Eq.(\ref{eq:6}) the TM reflection coefficient of graphene is given by
\begin{eqnarray}
r_{TM}^{(2)}=\frac{\frac{2e^2k_BT}{\pi\hbar^2}+\frac{e^2\xi_l^3}{8\pi k_BT}\int_0^{\infty}\frac{\tanh xdx}{x^2+(\hbar\xi_l /4k_BT)^2}}{\frac{\xi_l^2}{2\pi q_l}+\frac{2e^2k_BT}{\pi\hbar^2}+\frac{e^2\xi_l^3}{8\pi k_BT}\int_0^{\infty}\frac{\tanh xdx}{x^2+(\hbar\xi_l /4k_BT)^2}}.
\label{eq:14}
\end{eqnarray}

The equation suggests $r_{TM}^{(2)}(0,k_{\perp})=1$. In the same way, it can be found the expression of $r_{TE}^{(2)}(0,k_{\perp})$

\begin{eqnarray}
r_{TE}^{(2)}(0,k_{\perp})=-\frac{(2\pi /c)8\ln(2)\sigma_0}{\pi\lambda_Tk_{\perp}+(2\pi /c)8\ln(2)\sigma_0},
\label{eq:7}
\end{eqnarray}
where $\lambda_T=\hbar c/k_{B}T$ is the thermal wavelength in most materials. The explicit expression of $r_{TE}^{(2)}(0,k_{\perp})$ shows that it is susceptible to temperature.  $r_{TE}^{(2)}(0,k_{\perp})$ is nonzero only because of the intraband component of the graphene conductivity.

The first Matsubara frequencies $\xi_1$ are approximately 0.027, 0.162, and 0.325 $eV$ corresponding to the temperature 50, 300 and 600 $K$, respectively. As seen in Fig.~\ref{fig:6}, for $l \ge 2$, it is possible to substitute $\sigma_0$ for $\sigma(i\xi_l)$ in Eq.(\ref{eq:5}) to calculate the reflection coefficients and the higher order terms in the Casimir-Lifshitz force formula.

\section{RESULTS AND DISCUSSIONS}
In this section, we consider the Casimir interactions between graphene and a Fe metal plate. The parameters of Fe are $\omega_p=4.09$ $eV$, $\gamma_p=0.018$ $eV$, and $\mu(0)=10^4$ \cite{4}. As discussed in previous part $r_{TM}^{(1)}(0,k_{\perp})=r_{TM}^{(2)}(0,k_{\perp})=1$, the TM contribution to the $l=0$ term of the Casimir force is given

\begin{eqnarray}
F^{(0)}_{TM}(a,T)=-\frac{k_BT}{2\pi}\int_{0}^{\infty}\frac{k_{\perp}^2dk_{\perp}}{e^{2k_{\perp}a}-1}  =-\frac{k_BT\zeta(3)}{8\pi a^3}.
\label{eq:15}
\end{eqnarray}

The expression above is identical to the term with $l=0$ in the Casimir force between two metals. The sum of the $l \ge 1$ is much smaller than that of the metallic system due to the presence of the low graphene conductivity \cite{11}. As a result, the $l=0$ term is the dominating term. Now, to calculate the contribution of the TE mode $F^{(0)}_{TE}(a,T)$ with $l=0$, it is necessary to choose a good model for Fe. Both the plasma \cite{14} and Drude model \cite{22}, however, has been widely used to compute the Casimir interaction and fit with experimental data. Determining an accuracy of two models as compared to measurement data has been a controversial issue. In the following work, we utilize the plasma model to calculate the interactions. The expression of $F^{(0)}_{TE}(a,T)$ is expressed by
\begin{eqnarray}
F^{(0)}_{TE}(a,T)=-\frac{k_BT}{2\pi}\int_{0}^{\infty}\frac{k_{\perp}^2dk_{\perp}}{ \frac{e^{2k_{\perp}a}}{r_{TE}^{(1)}(0,k_{\perp})r_{TE}^{(2)}(0,k_{\perp})} -1}.
\label{eq:16}
\end{eqnarray}

\begin{figure}[htp]
\includegraphics[width=8.9cm]{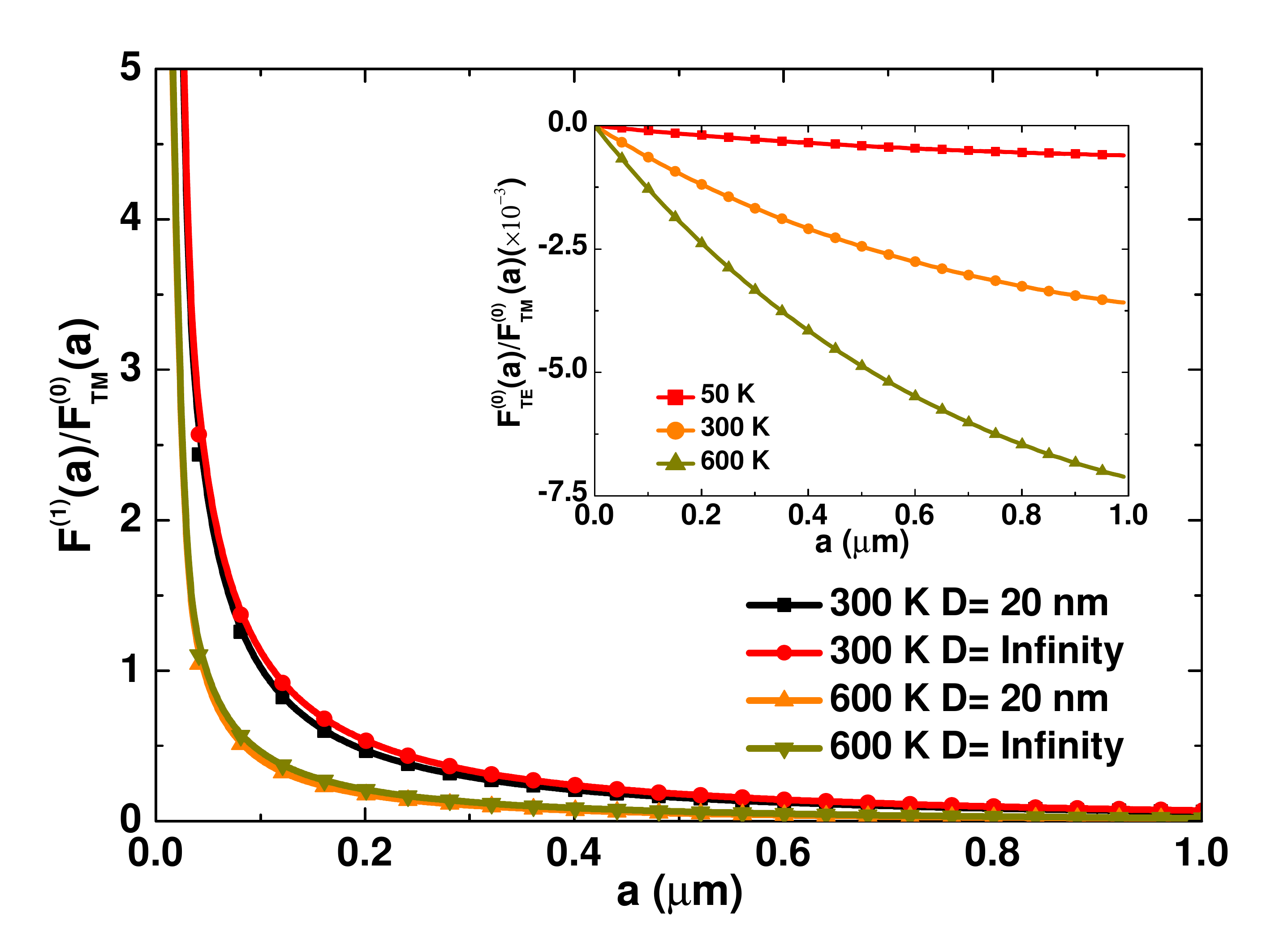}
\caption{\label{fig:7}(Color online) The ratios of $F^{(0)}_{TE}(a,T)/F^{(0)}_{TM}(a,T)$ and $F^{(2)}(a,T)/F^{(0)}_{TM}(a,T)$ at 50, 300, and 600 $K$ as a function of separation distance $a$.}
\end{figure}

Note that $F^{(0)}_{TE}(a,T)$ is a unique term affected by the magnetic property of the ferromagnetic substrate. In Fig.\ref{fig:7}, the ratio $F^{(0)}_{TE}/F^{(0)}_{TM}$ is less than 0.75 $\%$ for distances $a$ in the range from 0 to 1 $\mu m$ at given temperatures. It indicates that the effect of $\mu(0)$ on the Casimir force can be neglectable. For this reason, Fe can be treated as a regular non-magnetic metal $\mu(i\xi)=1$.

For a finite-thickness iron slab $D$, the reflection coefficients of the metal material are modified as follows \cite{15,23}

\begin{eqnarray}
R_{TE,TM}(i\xi_{l},k_{\perp})=r_{TE,TM}^{(1)}\frac{1-e^{-2k_1D}}{1-(r_{TE,TM}^{(1)})^2e^{-2k_1D}}.
\label{eq:9}
\end{eqnarray}

To study the influence of thickness on the Casimir force, we rewrite Eq.(\ref{eq:1}) in the following way
\begin{eqnarray}
F(a,T)=F^{(0)}_{TM}(a,T)+F^{(0)}_{TE}(a,T)+F^{(1)}(a,T),
\label{eq:17}
\end{eqnarray}
here $F^{(1)}(a,T)$ is the sum of all $l \ge 1$ terms and is thickness-dependent. Because $r_{TM}^{(1)}(0,k_{\perp})=1$, so $R_{TM}^{(1)}(0,k_{\perp})=1$. From this, the expression of $F^{(0)}_{TM}(a,T)$ is in Eq.(\ref{eq:15}) and is independent of $D$. The component force $F^{(0)}_{TE}(a,T)$ is weaker than that for the thick plate and one can consider $F^{(0)}_{TE}(a,T) \approx 0$. Therefore, only $F^{(1)}(a,T)$ is sensitive to a variation of subtrate thickness.

\begin{figure}[htp]
\includegraphics[width=9cm]{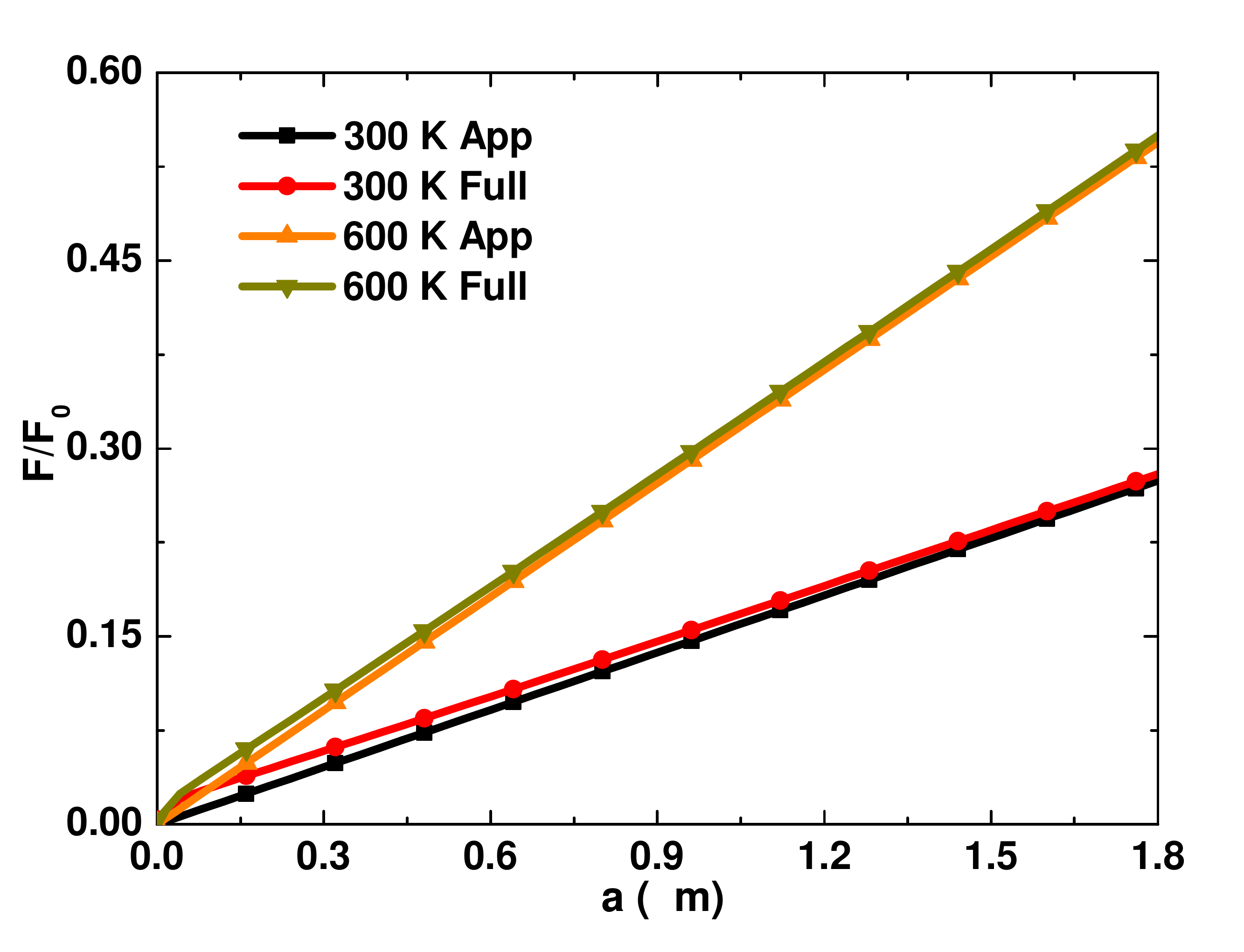}
\caption{\label{fig:2}(Color online) The Casimir pressures corresponding to the approximate and full expression at different temperatures. Here $F_0(a)=-\pi^2\hbar c/240a^4$.}
\end{figure}

Figure \ref{fig:7} shows that the thickness of the Fe slab does not influence much the Casimir forces. At the same temperature, the separation between the two curves corresponding to the finite plate and the semi-infinite plate is small. In bulk materials, a reduction of the thickness gives rise to a small decrease of the Casimir force. The thermal wavelength of graphene is $\lambda_T/200 \approx 38$ $nm$ at room temperature. At distances beyond the thermal wavelength of a system, the contribution of the sum of $l \ge 1$ terms can be ignorable. In the investigated range, the influence of $F^{(1)}(a,T)$ is not significant. This finding suggest that since graphene is a two dimensional monolayer material, it significantly interacts only with the nearest interface layers of the bottom substrate. Graphene's effect on layers separated by a sufficient number of layers is negligable. The Fe slab with $D = 20$ nm has more layers than that which are meaningful. Another interesting special feature is that at large distance limit ($a \ge 0.8 \mu m$) the contribution of $F^{(1)}(a,T)$ to the total Casimir force is minor. It can be explained that The $l \ge 2$ terms provide weak force components due to the presence of the small universal conductivity in the reflection coefficients. It is important to note that a universal graphene conductivity is a main reason for less than 2 $\%$ difference as one compares the dispersion force between two graphene sheets with $F_0(a)$ \cite{11}. It demonstrates that the approximate formula for the Casimir interaction equivalent to the $l=0$ term of the Casimir-Lifshitz expression in this regime is $F_{app}(a,T)=-k_BT\zeta(3)/8\pi a^3$. In Fig.\ref{fig:2}, we plot the normalized Casimir force with the full expression (see Eq.(\ref{eq:1})) and  the approximate expression. The $l =0$ term is dominant and can replace the full expression of the Casimir interaction when $a \ge 0.8$ $\mu m$ for 300 $K$.

The critical $T_c$ of Fe is 1043 $K$. At $T > T_c$, the magnetic properties of the ferromagnetic material nearly vanish. Spins in the Fe slab are rearranged so that $\mu(i\xi)=1$ at all frequencies. Nevertheless, the dielectric function is not affected by the directions of spins. Therefore, the plasma frequency and damping parameter are unchanged by temperature increases at $T \ge T_c$. If we consider the Casimir interaction between Fe and another metal, the phase transition leads to a significant change in the mutual interaction \cite{14,29}. The fact that graphene is quite transparent with the presence of the magnetic properties, the dispersion force may not be modified. 

A notable point is that the formula in Eq.(\ref{eq:6}) was used for pure graphene with the chemical potential $\mu_C = 0$. If $\mu_C \neq 0$, the graphene conductivity has to be rewritten in the form of \cite{25}
\begin{eqnarray}
\sigma(\mu_C,i\xi)=\frac{e^2k_{B}T\ln(2)}{\pi\hbar^2\xi}+\frac{e^2k_{B}T\ln(1+\cosh(\mu_C /k_{B}T))}{\pi\hbar^2\xi}\nonumber \\
+ \frac{e^2\xi}{\pi}\int_{0}^{\infty}\frac{\sinh(E/k_{B}T)}{\cosh(E/k_{B}T)+\cosh(\mu_C /k_{B}T)}\frac{dE}{(\hbar\xi)^2+4E^2}.\nonumber \\
\label{eq:11}
\end{eqnarray}

The first two terms are known as the intraband conductivity, another is the interband component. Eq.(\ref{eq:11}) shows that the intraband contribution of $\sigma(\mu_C,i\xi)$ still induces a non-zero value of $r_{TE}^{(2)}(0,k_{\perp})$. Obviously, the graphene conductivity is strongly susceptible to the chemical potential. The chemical potential of a graphene sheet can be controlled by using an applied electric field $E_d$ \cite{26,27}

\begin{eqnarray}
\frac{\pi\varepsilon_0\hbar^2v_F^2}{e}E_d=\int_{0}^{\infty}E\left(f(E)-f(E+2\mu_C)\right)dE,
\label{eq:13}
\end{eqnarray}  
here $f(E)$ is the Fermi distribution function, $v_F = c/300$ is the Fermi velocity.

Other ways to modulate $\mu_C$ are an applied magnetic field \cite{27} and chemical doping. Using a magnetic field also varys the expression of graphene conductivity and creates Landau energy levels. In the limit of extremely low chemical potential, $r_{TE}^{(2)}(0,k_{\perp})$ is represented as Eq.(\ref{eq:7}). For $\mu_C \gg k_{B}T$, the expression of the TE reflection coefficient with $l=0$ is written by
\begin{eqnarray}
r_{TE}^{(2)}(0,k_{\perp})\approx-\frac{4\sigma_0\mu_C/(\pi\hbar c)}{k_{\perp}+4\sigma_0\mu_C/(\pi\hbar c)}.
\label{eq:12}
\end{eqnarray}

\begin{figure}[htp]
\includegraphics[width=8.8cm]{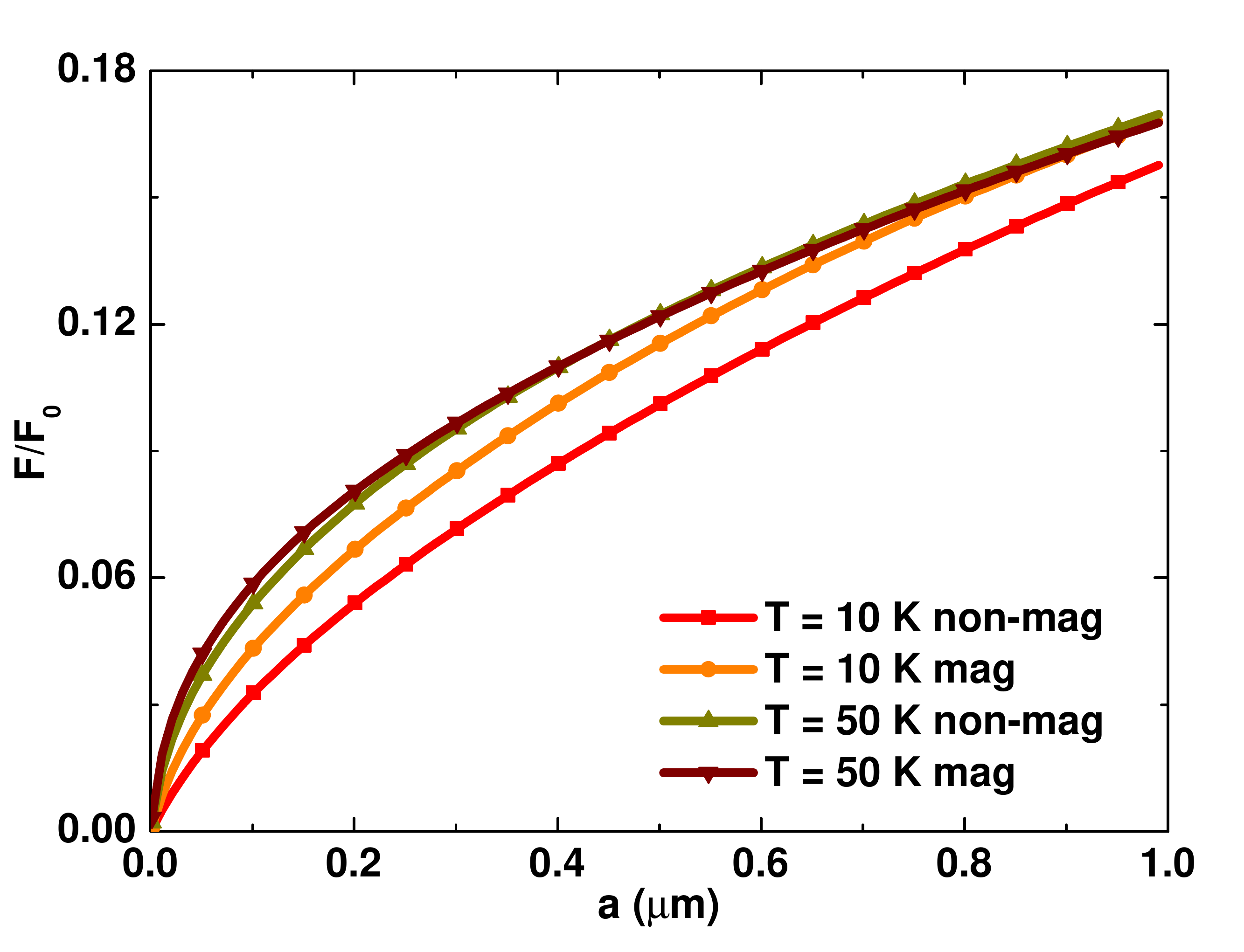}
\caption{\label{fig:3}(Color online) The Casimir forces at the case with and without taking into account the magnetic properties at $\mu_C = 1$ $eV$.}
\end{figure}

In Fig.\ref{fig:3}, the Casimir interactions are calculated at low temperatures ($\le 50$ $K$) and $\mu_C =1$ $eV$ with and without the magnetic properties. Unlike the case of a pristine graphene sheet ($\mu_C =0$ $eV$), the influence of permeability of Fe on the dispersion force is noticeable at 10 $K$. The properties, however, have much less effect on the Casimir force at temperatures greater than 50 $K$. The majority of $F(a,T)$ comes from the interband conductivity of graphene. At 10 $K$, the contribution of the intraband on the Casimir force can be considerable to that of the interband conductivity. However, as temperature increases the interband contribution increases substantially, and causing a decrease in the contribution from intraband. For this reason, the magnetic properties nearly disappear at $T \ge 50$ $K$.

\begin{figure}[htp]
\includegraphics[width=9.2cm]{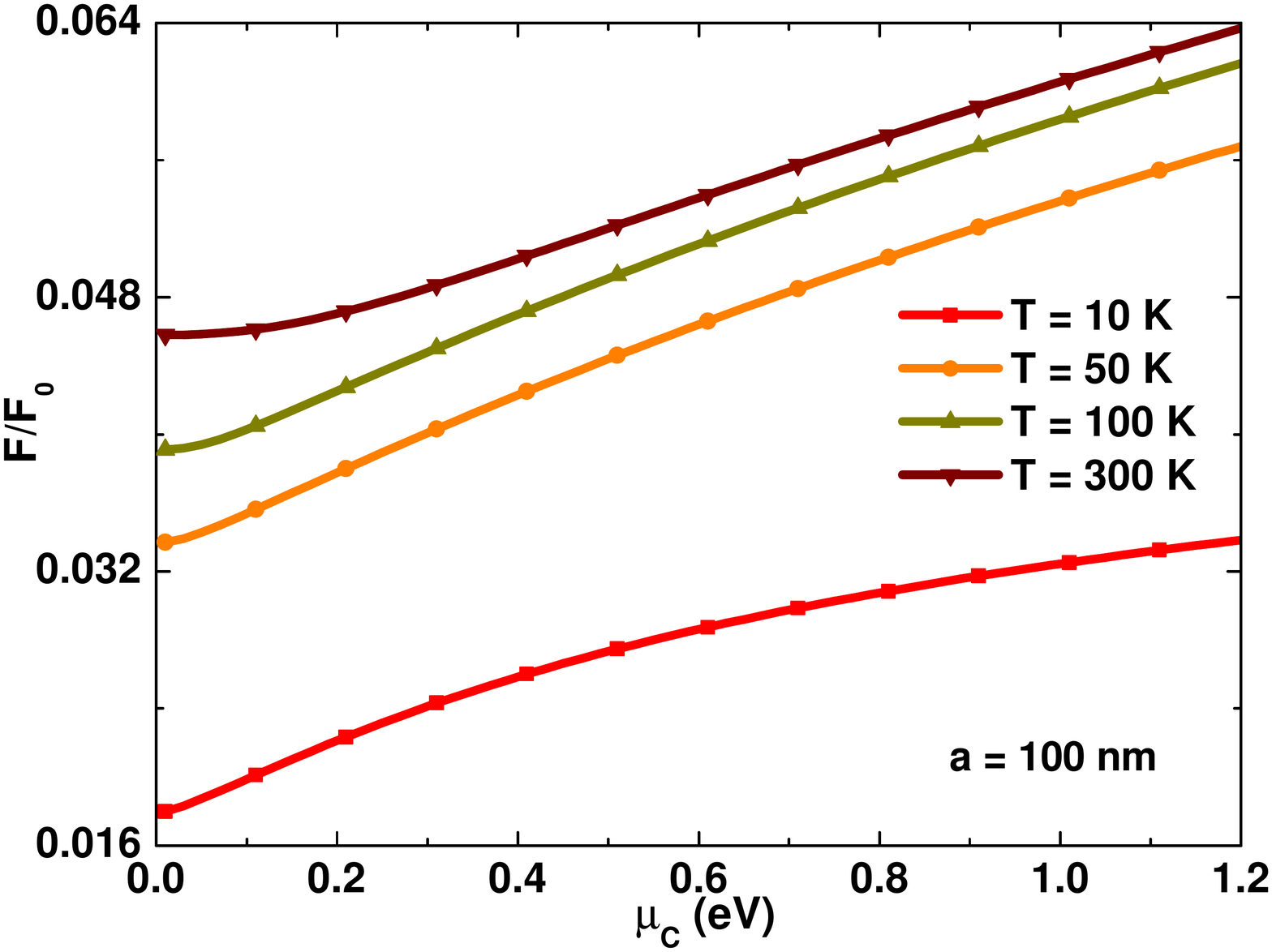}
\caption{\label{fig:4}(Color online) The Casimir pressures as a function of chemical potential.}
\end{figure}

To investigate further the Casimir force dependence on chemical potential, we consider the fluctuation interactions at $a = 100$ $nm$ and various temperatures versus $\mu_C$ shown in Fig.\ref{fig:4}. $F(a,T)/F_0$ changes from 0.018 to 0.034 at $T = 10$ $K$ and from 0.045 to 0.064 at $T = 300$ $K$ when $\mu_C$ varys in the regime below $1.2$ $eV$. In addition, at higher temperatures, the curves of $F(a,T)/F_0$ is of the linear form.

It is remarkable that Fe is not a superconducting material, so iron has no temperature phase transition as temperature decreases to 10 $K$. Alternatively, the universal value of conductivity $\sigma_0$ also demonstrates that even if temperature is 0 $K$, graphene still does not change phase and properties. 

\section{Conclusions}
When design and manufacturing of micro and nano devices and components, it is important and necessary to understand the loading and effects of related forces on the systems and its components, especially the nanoscale surface forces such as Casimir and the van der Waals. In this study, a comprehensive investigation and discussion of the Casimir interaction between graphene and FM materials is presented. We discussed about the Casimir force which is a function of the graphene conductivity and the permeability constant of $Fe$. It was shown that the conductivity of graphene heavily depends on the chemical potential which again has a large impact on the dispersion force and causes a difference between non-magnetic and magnetic calculations at the ultra-low temperature.  In other cases, there is no effect of the magnetic properties on the Casimir force. One can apply a bias electric field to a graphene sheet to tailor the chemical potential of graphene in order to control the magnitude of the Casimir force. According to our numerical results, the force between a graphene layer and the semi-infinite metal is smaller than the interaction between two metals in \cite{29} because the conductivity of metals is higher than that of graphene. The numerical results also demonstrated that the thickness of a metal slab had a minor influence on the Casimir force. The investigation and above findings about the Casimir force in this study would lead to useful information and effective solutions for design and manufacturing of micro and nano devices where graphene and FM metal materials are utilised, especially in the areas of micro and nano machining, fabrication, manipulation, assembly and metrology.

\begin{acknowledgments}
We give thanks to Dr. David Drosdoff for discussions. This research was supported by the Nafosted Grant No. 103.06-2011.51. Lilia M. Woods acknowledges the Department of Energy under contract DE-FG02-06ER46297. Nikolai A. Poklonski acknowledges the ﬁnancial support from Belarusian Republican Foundation for Fundamental Research Grant No. F11V-001.
\end{acknowledgments}

\end{document}